\documentclass[twocolumn,prd,unsortedaddress,superscriptaddress,showpacs,a4paper,nofootinbib]{revtex4-1}
\usepackage[T1]{fontenc}
\usepackage{graphicx}
\usepackage{indentfirst}
\usepackage{pdfpages}
\usepackage{multirow}
\usepackage{amssymb}
\usepackage{amsfonts}
\usepackage{amsmath}
\usepackage{epsfig}
\usepackage{soul}
\usepackage{slashed}
\usepackage{color}

\def\beq{\begin{equation}}
\def\enq{\end{equation}}
\def\beqa{\begin{eqnarray}}
\def\enqa{\end{eqnarray}}
\def\MeV{\nobreak\,\mbox{MeV}}
\def\GeV{\nobreak\,\mbox{GeV}}

\def\mix{\lag\bar{q}g\si.Gq\rag}

\def\G3{\lag g^3G^3\rag}
\def\pli{p^\prime}

\def\la{\lambda}

\def\Ga{\Gamma}

\def\si{\sigma}

\def\al{\alpha}

\def\lb{\label}
\def\nn{\nonumber}
\newcommand{\rag}{\rangle}
\newcommand{\lag}{\langle}

\begin{document}
\title{A QCD sum rule calculation  of the  $X^\pm(5568) \to B_{s}^0\pi^\pm$ decay width}

\author{ J. M. Dias}
\affiliation{Instituto de F\'{\i}sica, Universidade de S\~{a}o Paulo, 
C.P. 66318, 05389-970 S\~{a}o Paulo, SP, Brazil}

\author{K. P. Khemchandani}
\affiliation{Faculdade de Tecnologia, Universidade do Estado do Rio de 
Janeiro, Rod. Presidente Dutra Km 298, P\'olo Industrial, 27537-000 , 
Resende, RJ, Brasil}

\author{A. Mart\'inez Torres}
\affiliation{Instituto de F\'{\i}sica, Universidade de S\~{a}o Paulo, 
C.P. 66318, 05389-970 S\~{a}o Paulo, SP, Brazil}

\author{M. Nielsen}
\affiliation{Instituto de F\'{\i}sica, Universidade de S\~{a}o Paulo, 
C.P. 66318, 05389-970 S\~{a}o Paulo, SP, Brazil}

\author{C. M. Zanetti}
\affiliation{Faculdade de Tecnologia, Universidade do Estado do Rio de 
Janeiro, Rod. Presidente Dutra Km 298, P\'olo Industrial, 27537-000 , 
Resende, RJ, Brasil}

\begin{abstract}
To understand the nature of the $X(5568)$, recently observed in
the  mass spectrum of the $B_{s}^0\pi^\pm$ system by the D0
Collaboration, we have investigated, in a previous work, a scalar tetraquark
(diquak-antidiquark) structure for it, within the two-point 
QCD sum rules method. We found that it is possible to obtain a stable 
value of the mass compatible with the D0 result, although a rigorous QCD 
sum rule constrained analysis led to a higher value of mass. As a 
continuation of our investigation, we calculate the width of the tetraquark 
state with same quark content as $X(5568)$, to the channel $B_{s}^0\pi^\pm$, 
using the three-point QCD sum rule. We obtain a value of $(20.4\pm8.7)\MeV$ 
for the mass $\sim$ 5568 MeV, which is compatible with the experimental 
value of $21.9\pm6.4(\mbox{sta})^{+5.0}_{-2.5}(\mbox{syst}) \MeV/c^2$. 
We find that the decay width to $B_{s}^0\pi^\pm$ does not 
alter much for a higher mass state.
\end{abstract} 


\maketitle


The  D0  Collaboration   has  recently  reported  the   study  of  the
$B_{s}^0\pi^\pm$ mass spectrum in  the energy range 5.5-5.9~GeV, where
a narrow enhancement of the experimental data is found and interpreted
as a  new state: $X(5568)$~\cite{D0:2016mwd}.  The mass and  width for
this    state    have    been     found    to    be    $m=5567.8\pm2.9
(\mbox{sta})^{+0.9}_{-1.9}(\mbox{syst})          \MeV/c^2$         and
$\Gamma=21.9\pm6.4  (\mbox{sta})^{+5.0}_{-2.5}(\mbox{syst}) \MeV/c^2$,
respectively \cite{D0:2016mwd}.  The isospin  of $X(5568)$  is clearly
one. Its  spin-parity is not  yet known  although a scalar  four quark
interpretation has been suggested in Ref.~\cite{D0:2016mwd}.

The finding of this new state adds to the rigor with which the exotic hadrons 
with heavy quark flavor are being studied currently.  Until just about a 
decade ago, the data related to the spectroscopy of hadrons with open or 
hidden charm/bottom structure were relatively scarce and of poor statistical 
quality.  However, the scenario has changed rapidly during the last few years 
with the working of new experimental facilities like LHCb, BES, BELLE, etc., 
and good quality experimental data is being published continuously. With 
sufficient amount of data available, it has been possible to identify several 
new states, actually way too many to fit in the traditional quark-antiquark 
spectrum. Indeed, theoretical studies indicate that many of these hadrons must 
be exotic in nature. For example, the first such state discovered in 
the charm sector is the $X(3872)$ \cite{pdg}. The mass as well as the narrow 
width  of $X(3872)$, $\Gamma < 1.2$ MeV \cite{pdg}, inspite of  having a large 
phase space for decay to some open channels,  cannot be explained within the 
conventional quark model. A series of similar states have been found and 
their structure, quantum numbers, etc., are being debated continuously in 
the literature. Recently, even clearer evidence of the exotic nature has 
been brought forward with the finding of special mesons, which are heavy 
quarkonium-like but at the same time are electrically charged. Such states 
would at least require four valence quarks to get the nonzero charge. 
Some examples of such charged charmonium-like states are: $Z_c(3900)$, 
$Z_c(4025)$, $Z_c(4250)$, $Z_c(4430)$ in the charm sector \cite{zc1,zc2,zc3,zc4} 
and $Z_b(10610)$, $Z_b(10650)$ \cite{zb} in the bottom sector. The $X(5568)$ is an addition to 
the list of undoubtedly exotic mesons, since its wave function consists of 
four different flavors: $u$, $b$, $d$ and $s$ quark.

The observation  of this  new  state  has already motivated  several  theoretical
investigations \cite{Agaev, us, Wang, Chen, WangZhu, LiuLiuZhu, LiuLi,
XiaoChen, Agaev2}. In Refs.~\cite{Agaev,us, Wang, Chen} the 
calculations for  the mass  of $X(5568)$  have been  done using the  
QCD sum rules  (QCDSR) method, and results  in excellent agreement  
with the experimental  value have been found.   
In Refs.~\cite{Agaev,us, Wang} $J^{PC} =  0^{++}$ 
was assumed while in Ref.~\cite{Chen} scalar as
well as  axial tetraquark currents were considered.   In
Ref.~\cite{WangZhu}  a model  using multiquark  interactions has  been
used and  a 150 MeV higher  mass is found for  $X(5568)$, although the
systematic   errors  still   allow  their   state  to   be  related   to
$X(5568)$. Another multiquark  model calculations using color-magnetic
interaction has been presented in \cite{LiuLiuZhu}. The possibility of
explaining the enhancement in the  data as near threshold rescattering
effects has been  studied in Ref.~\cite{LiuLi}.  The $B \bar  K$ and $B^*
\bar   K$   molecular   interpretations   have   been   suggested   in
Ref~\cite{XiaoChen}. A calculation of the  width of $X(5568)$ has also
been reported  in Ref.~\cite{Agaev2}  using sum  rules based  on light
cone  QCD,  but  with  a  Lagrangian  that  is  not  usual,  with
derivatives in the heavy particle fields and not in the pion field.

In Ref.~\cite{us} we investigated if the mass of $X(5568)$ can 
be reproduced in terms of a scalar diquark-antidiquark current (an 
isovector analog of the $D_{s0}^{\pm}(2317)$ description presented 
in Ref.~\cite{nos}). We found that a stable value of mass can be 
obtained around 5568 MeV while ensuring the dominant 
contribution to come from the pole. However, further analysis 
showed that requiring a simultaneous convergence of the operator 
product expansion series on the QCD side leads to a higher value of 
the mass, $\sim$ 6390 MeV. 

As a continuation of our investigation, 
we  now calculate the decay width of the state found in Ref.~\cite{us}
to $B_{s}^{0}\pi^\pm$, following the calculations of the analogous  
decay  in the  charm  sector done in Ref.~\cite{nielsen}. Before proceeding 
further, it is important to note that for a state with mass $\sim$ 5568 MeV, 
the decay channels $B  \bar K$, $B^* \bar K^*$ and $B_{s}^{*0}\rho$
are closed. Moreover the $0^{++}$ spin-parity assignment would not 
allow the decay to the other possible open channel $B_{s}^{*0}\pi^\pm$. 
Even the radiative decay will be allowed only for the neutral 
member of the isospin triplet. In this case, the decay width to
$B_{s}^{0}\pi^\pm$ should be comparable to the experimental total width 
\cite{D0:2016mwd}. However, for a higher mass, the decay to other
channels may contribute to the total width. We calculate the width in 
both cases and present more discussions related to the results obtained.
For the sake of convenience, we shall refer to our state as $X^\pm$ in the 
following discussions.

In Ref.~\cite{us} we considered a scalar diquark-antidiquark current 
in terms of the interpolating field:
\beqa
j_{X}&=&{\epsilon_{abc}\epsilon_{dec}}(u_a^TC
\gamma_5s_b)(\bar{d}_d\gamma_5C\bar{b}_e^T),
\label{int}
\enqa
where $a,~b,~c,~...$ are colour indices, $C$ is the charge conjugation
matrix. 

To calculate the vertex, $X^+  B_s^0 \pi^+$, we use
 the three-point correlation function 
\begin{align}\nonumber
&\Gamma_\mu(p,\pli,q)=\\&\int d^4x \int d^4y e^{i\pli.x} e^{iq.y}
\lag 0 |T[j_{B_{s}^0}(x)j_{5\mu}^{\pi}(y)j^\dagger_{X}(0)]|0\rag,
\lb{3po}
\end{align}
where $p=\pli+q$ and the interpolating currents for the pion and $B_s^0$ mesons 
are given by:
\begin{align}\nonumber
j_{5\mu}^\pi& = \bar{d}_a\gamma_\mu\gamma_5 u_a, \,\,
\\ j_{B_s^0}& = i\bar{b}_a\gamma_5 s_a.
\lb{pseu}
\end{align}

As the standard procedure in the QCDSR calculations
\cite{Nielsen:2009uh,Nielsen:2014mva}, we use the dual interpretation
of the correlation function and assume  that there is an interval over
which Eq.~(\ref{3po})  may be equivalently described at the quark as
well as the hadron level. Following this
assumption:  
\begin{enumerate}
\item{   On the OPE side, the vertex function of
    Eq.~(\ref{3po}) is calculated in terms of quark and gluon fields
    using the Wilson's operator product expansion (OPE).}
\item{ On the phenomenological side, the
    same function is then calculated by treating the currents as
    the creation and annihilation operators of hadrons and as a result
    hadron properties, such as masses and coupling constants, are
    introduced in the process.}
\item{Finally, both results are equated to extract the value of the coupling 
constant required to obtain the width of the state.}
\end{enumerate}

The phenomenological side is calculated by 
inserting intermediate states for $B_{s}^0$, $\pi^+$ and $X^+$ in Eq.~(\ref{3po}), and by 
using the definitions: 
\beqa
&&\lag 0 | j_{5\mu}^{\pi^0}|\pi(q)\rag =iq_\mu F_{\pi},\;\;\;\;\\
&&\lag 0 | j_{B_{s}^0}|B_{s}^0(\pli)\rag ={m_{B_{s}^0}^2f_{B_{s}^0}\over 
m_b+m_s},\;\;\;\;\\
&&\lag 0 | j_{X}|X(p)\rag =\la_X,
\lb{fp}
\enqa
we obtain the following relation:
\beqa
\Ga_{\mu}^{phen} (p,\pli,q)&&=\frac{\la_X m_{B_{s}^0}^2f_{B_{s}^0}F_{\pi}\,g_{XB_{s}^0\pi}\,\,q_\mu}{(m_b+m_s)(p^2-m_{X}^2)({\pli}^2-m_{B_{s}^0}^2)(q^2-m_\pi^2)}\nonumber\\&& +\,\mbox{continuum contribution},
\lb{phen}
\enqa
where the coupling constant $g_{XB_{s}^0\pi}$ is defined by the on-mass-shell
matrix element,
\beq
\lag B_{s}^0 \pi|X\rag=g_{XB_{s}^0\pi}.
\label{coup}
\enq

The second term on the right-hand side in Eq.~(\ref{phen}) contains the contributions of
all possible excited states.

We follow Refs.~\cite{nielsen,z3900} and work at the pion pole, as suggested in 
\cite{rry} for the pion-nucleon coupling constant. We do this because the matrix element in Eq.~(\ref{coup})
defines the coupling constant only at the pion pole. For $q^2\neq0$ one would 
have to replace de coupling constant $g_{XB_s^0\pi}$, in Eq.~(\ref{coup}), 
by the form factor $g_{XB_{s}^0\pi}(q^2)$ and, therefore, one would 
have to deal with the 
complications associated with the extrapolation of the form factor 
\cite{bclnn,Bracco:2011pg}. The pion pole method consists in neglecting the pion 
mass in the denominator of Eq.~(\ref{phen}) and working at $q^2=0$. On the 
OPE side one singles out the leading terms in the operator product expansion 
of Eq.(\ref{3po}) that match the $1/q^2$ term. On the other hand, from phenomenolo\-gi\-cal side, 
we get the following expression for the ${q_\mu / q^2}$ structure,

\beqa
\Ga^{phen}(p^2,{\pli}^2)&&=\frac{\la_X m_{B_{s0}}^2f_{B_{s0}}F_{\pi}\,g_{XB_{s0}\pi}}{(m_b+m_s)(p^2-m_{X}^2)({\pli}^2-m_{B_{s}^0}^2)}\nonumber\\
&&+\int_{m_b^2}^\infty{\rho_{cont}(p^2,u)\over u-{\pli}^2}~du.
\label{mco}
\enqa

In  Eq.~(\ref{mco}), $\rho_{cont}(p^2,u)$, gives the continuum
contributions, which can be parametrized as
$\rho_{cont}(p^2,u)={b(u)\over s_0-p^2}\Theta(u-u_0)$ 
\cite{io2}, with $s_0$ and  $u_0$ being the continuum thresholds 
for $X^+$ and $B_{s}^0$, respectively. Since 
we are working at  $q^2=0$, we  take the limit  $p^2={\pli}^2$ and we 
apply the Borel transformation to $p^2\rightarrow M^2$ and get:
\beqa
\Ga^{phen}(M^2)&&= \frac{\la_X m_{B_{s}^0}^2f_{B_{s}^0}F_{\pi}~
g_{XB_{s0}\pi}}{(m_b+m_s)(m_{X}^2-m_{B_{s}^0}^2)}\left(
e^{-\frac{m_{B_{s}^0}^2}{M^2}} -e^{-\frac{m_{X}^2}{M^2}}\right)\nonumber\\&&+A~e^{-\frac{s_0}{M^2}}+
\int_{u_0}^\infty\rho_{cc}(u)~e^{-u/M^2}du,
\label{paco}
\enqa
where A is a parameter introduced to take into account 
pole-continuum transitions, which are not supressed when only a 
single Borel transformation is done in a three-point function sum rule \cite{io2,eidemuller}. 
For simplicity, one assumes that the pure continuum contribution to the 
spectral density, $\rho_{cc}(u)$, is related to the spectral density obtained on the OPE 
side, $\rho_{OPE}(u)$, through the ansatz: $\rho_{cc}(u)=\rho_{OPE}(u)\Theta(u-u_0)$. 

As  discussed  in Refs.~\cite{navarra2006,z3900},  large  partial
decay widths are expected when  the coupling constant is obtained from
QCDSR  in the  case of  multiquark states, that contains 
the same valence quarks as the valence quarks in the final state.  This happens  
because, although the initial current,  Eq.~(\ref{int}), has a non-trivial color 
structure, it can be rewritten as a sum of molecular type currents with 
trivial color configuration through a Fierz transformation, as explicitly 
shown in Ref.~\cite{Chen}. To avoid this problem we follow 
Refs.~\cite{navarra2006,z3900}, and consider on the OPE side only diagrams 
with non-trivial color structure, which are called color-connected (CC) diagrams.

On the OPE side we compute the CC diagrams working at leading order in
$\al_s$. Singling out the leading terms proportional to $q_\mu/q^2$, up to 
dimension five the only diagrams that contribute are proportional
to the mixed condensate.  We
can write the  Borel transformation of the correlation function on the OPE
side in terms of a dispersion relation:
\beq
\Ga^{OPE}(M^2)=\int_{m_b^2}^\infty  \rho_{OPE}(u)~e^{- u/M^2}du\;,
\lb{ope}
\enq
where the spectral density, $\rho_{OPE}$, is given by the imaginary part of 
the correlation function. Transferring the pure continuum contribution to
the OPE side, the sum rule for the coupling constant is 
given by:
\beq
C~\left(e^{-m^2_{B_{s}^{0}}/M^2} - 
e^{-m_{X}^2/M^2}\right)+A~e^{-s_0/M^2}=\nn
\enq
\beqa
\frac{\mix}{2^5\pi^2}\int_{m_b^2}^{u_0}\left(\frac{2}{3}-\frac{m_b^2}{u}\right)e^{-u/M^2},
\label{sr}
\enqa
with
\beq
C={\la_X m_{B_s^0}^2f_{B_s^0}F_{\pi}~g_{XB_s^0\pi}
\over(m_b+m_s)(m_{X}^2-m_{B_s^0}^2)}.
\label{C}
\enq

The values  of the phenomenological  parameters used in  the numerical
analysis   of   the  sum   rules are   the same as used in Ref.~\cite{us} and are
listed  in   Table~\ref{tab1}.  The  meson-current  coupling  
$\la_X$  is obtained from the two-point correlation funcion \cite{us}.

\begin{center}
\begin{table}[h]
\begin{tabular}{p{3.5cm} p{3.5cm}}
\hline
Parameters & Values \\
\hline
 $m_s$ &  $(0.13\pm0.03)\,\GeV$ \\
 $m_b$ & $(4.24\pm0.06)\,\GeV$ \\
 $m_{B_S^0}$  &  $5.366\GeV$                \\
 $F_\pi$ &  $93\sqrt{2}\MeV$          \\ 
 $f_{B_S^0}$  &   $(0.224\pm0.014)\GeV$              \\
 $\la_X$    & $(9.36\pm1.38) \times 10^{-3}\GeV^5$ \\    
 $\lag\bar{q}q\rag$ &  $-(0.23\pm0.03)^3\,\GeV^3$ \\
 $m_0^2=\lag\bar{q}g\si.Gq\rag/\lag\bar{q}q\rag$  &  $0.8\,\GeV^2$ \\
 \hline
\end{tabular}\caption{QCD input parameters \cite{us,SNB,pdg,Bernardoni:2014fva}.}
\label{tab1}\end{table}
\end{center}


\begin{figure} \label{fig1}
\centerline{\epsfig{figure=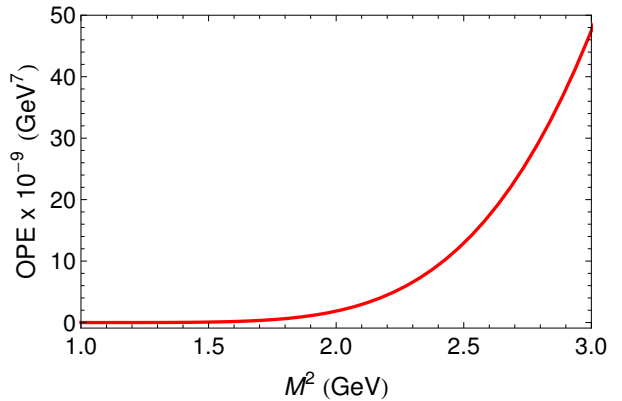,width=0.44\textwidth}}
\caption{The OPE side   (RHS of Eq.(\ref{sr})), as a function of the Borel mass.} 
\end{figure} 

\begin{figure}
\centerline{\epsfig{figure=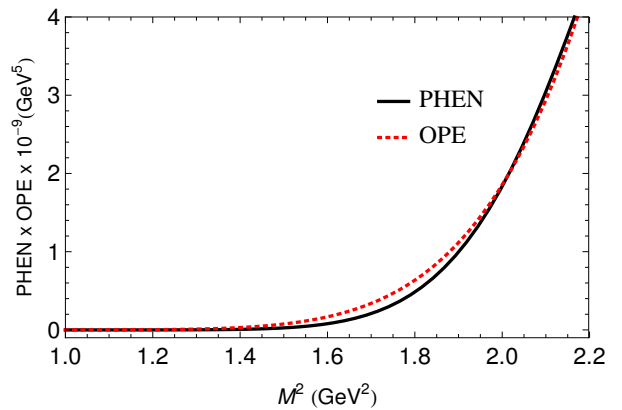,width=0.44\textwidth}}
\caption{The phenomenological  side (solid) and the  OPE side (dashed)
  of Eq.(\ref{sr}), as a function of the Borel mass.} \label{fig1}
\end{figure}

In Fig.~1 we show  the OPE side of the sum rule (right-hand side
of Eq.~(\ref{sr})), for $\sqrt{s_0}=6.0\GeV$ and $\sqrt{u_0}=5.866\GeV$, as 
a function  of the  Borel mass. The value of $\sqrt{s_0}=6.0\GeV$
is within the range used in Ref.~\cite{us} which resulted in a mass: $m = 5.58 \pm 17$ 
GeV, compatible with the $X^+(5568)$ mass. We  can see  that the  OPE shows  good
stability up to  $M^2\le2.2\GeV^2$.  Therefore, in our  analysis we will
consider  the window  for the  Borel mass  in the  range $1.0\GeV^2\le
M^2\le2.2\GeV^2$.

In order to  determine the coupling constant, 
$g_{XB_{s}^0\pi}$, we fit the QCDSR results with 
the analytical expression on the left-hand side
(the phenomenological side) of Eq.~(\ref{sr}) within the 
determined Borel window. In Fig.~2 we show the fit 
obtained for the phenomenological side, as well as 
the OPE results, for $\sqrt{s_0}=6.0\GeV$, and
$\la_X=9.236\times~10^{-3}\GeV^5$  (which is the 
value for the same $s_0$). From Fig.~2 we can see that 
the fit reproduces well the OPE results. The stability of the 
of the OPE results guarantees that the values of the fit 
parameters do not vary with the Borel mass. 
The fit requires determining two unknown parameters: 
$C$ which is related to the coupling constant $g_{XB_{s}^0\pi}$ 
and the weight $A$ from the pole-continuum transitions. Only 
the former is useful for us in the present calculation since 
with the value obtained for the parameter $C$ from the fit, we 
can extract the value of the coupling constant using 
Eq.~(\ref{C}) and the parameters in Table~\ref{tab1}. The value 
obtained for the coupling constant is $g_{XB_S^0\pi}=10.3~\GeV$. 

The result obtained for the coupling shows a large 
dependence on the parameters  $s_0$ and $\lambda_X$.  To 
determine the  error introduced  by these parameters, we allow 
$s_0$ to vary    in    the    interval $5.9\leq\sqrt{s_0}\leq6.1~\GeV$ 
(the same interval used in the study of the mass in Ref.~\cite{us}),  
and the  error  for the  $\lambda_X$ shown in Table.~1. Considering 
the uncertainties given above, we finally find:

\beq
g_{XB_S^0\pi}=(10.3\pm2.3)\GeV\label{coupling} \, .
\enq

It is important to point out that this coupling has the same dimension as the 
usual coupling between one scalar and two pseudoscalar mesons \cite{sca}.
This is not the case of the $XB_s\pi$ coupling determined in Ref.~\cite{Agaev2},
which is given in units of $\GeV^{-1}$, since the authors use a higher dimension
interaction Lagrangian,  with two derivatives in the heavy particle fields.

The coupling constant, $g_{XB_s^0\pi}$, 
is related with the partial decay width as:
\beqa
\Gamma(X(5568)\rightarrow B_s^0\pi^+)=\frac{g_{XB_s^0\pi}^2}{16\pi m_{X}^3}
\sqrt{\la(m_{X}^2,m_{B_s^0}^2,m_{\pi}^2)},\nn\\
\lb{decay}
\enqa
where $\la(a,b,c)=a^2+b^2+c^2-2ab-2ac-2bc$. Considering the 
result for the coupling constant, Eq.~(\ref{coupling}), we obtain 
the width of the decay:
\beq
\Gamma(X^+(5568)\rightarrow B_s^0\pi^+)=(20.4\pm8.7)\MeV\;.
\label{fin}
\enq 

This result is  in good agreement with the experimental 
total decay width~\cite{D0:2016mwd}: 
\beq
\Gamma^{exp}=21.9\pm6.4  (\mbox{sta})^{+5.0}_{-2.5}(\mbox{syst}) \MeV\;,
\enq
which is expected since, as commented above, the 
$B_s\pi$ channel is the dominant decay channel for 
the charged $X^\pm(5568)$ state. In Ref.~\cite{Agaev2} the 
authors find $\Gamma(X^+(5568)\rightarrow B_s^0\pi^+)=(24.5\pm8.2)\MeV$, 
which is also in good agreement with the experimental value.

Finally, we recall that a more rigorous analysis, requiring 
the constraints of the dominance of the pole contribution 
on the phenomenological side of QCDSR calculations together
with the convergence of the OPE series on the QCD side, led 
to a higher value of mass: $(6.39\pm0.10)$ GeV \cite{us}. Such 
a value is not in agreement with the one found by the D0 
Collaboration \cite{D0:2016mwd}. However, more recently the 
LHCb Collaboration has not confirmed the observation of the X(5568) 
and no structure is found in their $B_{s}^0\pi^\pm$ mass 
spectrum from the  threshold up to $\sim 5700\GeV$. More 
analyses are required to clarify this situation. We have also evaluated 
the coupling and the width corresponding to this higher value of the 
mass. We find that Borel stability and the phenomenological versus 
OPE agreement are comparable with the ones presented in Figs.~1 
and 2. Using the values obtained in \cite{us} for $\lambda_X$ and 
$s_0$: $\lambda_X=4.75\times10^{-2}~\GeV^5$, $s_0=(48\pm2)~\GeV^2$, 
the value of the coupling $g_{XB_s^0\pi}$ for the mass $(6.39\pm0.10)$ 
GeV turns out to be 
\beq
g_{XB_s^0\pi}=(5.7\pm 0.8)~\GeV,
\label{coupling2}
\enq
and consequently the width obtained is 
\beq
\Gamma(X^+\rightarrow B_s^0\pi^+)=(30.1\pm 8.6) \MeV.
\enq 
We can see that the result of the width has not altered much, 
although more phase space is available for decay with mass 6.39 GeV.  The 
reason for that is because the coupling of the  heavier state is
weaker to the $B_{s}^0\pi^\pm$ channel, as can be seen by 
comparing Eqs.~(\ref{coupling}) and (\ref{coupling2}). Further, it 
should be noticed that the width now is only a partial width and 
contributions from other decay channels like, $B \bar K$ , $B^* 
\bar K^*$, $B^{*0}_s \rho$, may contribute to the total value. 
Investigations  can be made in this direction in future.

In  summary, we have presented a QCD sum rule study of the vertex
function associated with the strong decay $X^+\rightarrow
B_s^0\pi^+$, using a three-point function QCD sum rule approach. 
Following the study presented in \cite{us}, the $X$ meson was 
considered as a scalar diquark-antidiquark state. To extract
directly the coupling constant from the sum rule we work at the pion pole
and we consider only color connected diagrams to ensure the non-trivial color 
structure of the tetraquark current. The value obtained for the 
$g_{XB_s^0\pi}$ coupling was used to evaluate the decay width 
for this channel. In Ref.~\cite{us} it was shown that a stable mass 
compatible with $X(5568)$ can be obtained although a more 
constrained analysis results in a higher value of the mass. We have calculated the 
width in both cases and find that result does not alter much with the mass.

\vspace{1cm}

\underline{Acknowledgements}: 
This work has been supported by CNPq and FAPESP.

\end{document}